\documentclass[12pt]{iopart}
\usepackage{iopams}
\begin{document}

\title[Bulk scalar field in DGP braneworld cosmology ]{Bulk scalar field in DGP braneworld cosmology }

\author{Rizwan ul Haq Ansari and P K Suresh}

\address{School of Physics, University of Hyderabad, Hyderabad 500 046. India.}
\ead{ph03ph14@uohyd.ernet.in; pkssp@uohyd.ernet.in}
\begin{abstract}
 We investigated the effects of bulk scalar field in the braneworld cosmological scenario. The Friedmann equations and acceleration condition in  presence of the  bulk scalar field  for a zero tension brane  and cosmological constant are studied. In  DGP  model the  effective Einstein equation on the brane is obtained with bulk scalar field. The rescaled bulk scalar field  on the brane in
the DGP model behaves as an effective four dimensional field,  thus standard type cosmology is  recovered. In present study of the  DGP  model, the late-time accelerating phase of the  universe can be explained .
\end{abstract}



\maketitle

\section{Introduction}
In recent times theories with higher-dimensions have received much attention  in high
 energy physics, specially in  the context of hierarchy problem, and cosmology  \cite{horava}.
 In this scenario it is very likely that our  four dimensional universe is a subspace called brane embedded in higher
 dimensional space-time called bulk. As a realization of such higher-dimensional theory, the brane world scenario has
 attracted a lot of attention in cosmology. It is believed that the new scenario has the potential to address several
 issues in  cosmology  like dark energy, current accelerating phase of the universe, cosmological constant
  and inflation.

In 2000, Dvali, Gabadadze and Porrati (DGP) introduced a braneworld
model of gravity \cite{dvali}. In this model our universe is a brane
embedded in bulk and the standard model particles are confined on
the brane and gravity propagates in the bulk ( for a review  of DGP
model see \cite{lue}). In DGP braneworld, gravity is modified at
large distance rather than at short distance in contrast to other
popular braneworld scenarios, because of an induced four-dimensional
Ricci scalar in the action. The DGP cosmological  model posses a
solution which is  "self-accelerating" phase.
 In the studies of  DGP model so far, bulk is considered as empty except for a cosmological constant and the
 matter fields on the brane are considered as responsible for the evolution on  the brane.

Braneworld models with non-empty bulk or scalar field in the bulk
has been discussed by various authors \cite{maria,himem}.
It is  believed that in the unified theory approach, a dilatonic gravitational
 scalar field term is required in the the five dimensional Einstein-Hilbert action \cite{maeda}.
 One of the first motivations to introduce a bulk
scalar field is  to stabilize the distance between the two
branes \cite{olive} in the context of the first model introduced by
Randall and Sundrum. A second, motivation for studying scalar fields
in the bulk is due the possibility that such a setup could provide some
clue to solve the famous cosmological constant problem. Models with
inflation driven by bulk scalar field have been studied and it is shown
that inflation is possible without inflaton on the brane
\cite{himem}. Later the quantum fluctuations of brane-inflation and
reheating issues are  also addressed in the braneworld scenario with
bulk scalar field \cite{yoko}. The creation of a brane world
with a bulk scalar field using an instanton solution in a five dimensional
Euclidean Einstein equations is also considered \cite{ayan}.

In the present study, we consider the effects of scalar field living in the bulk, on the dynamics of
 brane. The effective Einstein equations on the brane  with a zero tension  and  its consequence in the braneworld cosmology  can be studied. It is possible to show that such a model with  bulk scalar field can be used to obtain
the standard Friedmann type of cosmology  without introducing a tension or cosmological constant. The
acceleration conditions  for a universe dominated with bulk scalar field  is also  worth to be examined.
The DGP model also has  zero tension brane and infinite size extra-dimension, so we are also interested
 in  the DGP model with  bulk-scalar field  and corresponding cosmological solutions. The Friedmann
 type cosmology can be obtained in the DGP braneworld cosmology, under suitable conditions, without matter
 field on the brane. It can be shown that, in our current frame work of  the  DGP braneworld, the solution
of Einstein equation exhibit a  self accelerating phase of the universe. The rescaled bulk scalar field on
the brane in the DGP model is expected to  mimic the inflaton on the brane and
thus standard cosmology could be recovered.

\section{Effective Einstein equation for tension free brane }
Braneworld  models without/with  a  cosmological constant \cite{langlois, binetury} and infinite
size flat extra dimensions have been received with much attention
recently \cite{gumj}. Consider a three dimensional
brane embedded in a five dimensional bulk and  assume that the five
 dimensional metric  have the following form
\begin{equation} \label{met}
ds^{2}= g_{AB}dx^{A}dx^{B} = dy^2 + q_{\mu\nu}dx^{\mu}dx^{\nu}.
\end{equation}
Here onwards the Latin indices are running from 0 to 4 and Greek indices
are from 0 to 3.

The Einstein equation for the metric (1) with source  can be written
as ( with $k_5^2=8\pi G_5$ )
\begin{equation}
R_{AB}-\frac{1}{2}g_{AB}R= -\Lambda g_{AB}+ k_5^2 (T_{AB} + S_{AB} \delta(y)).
\end{equation}
Where  $T_{AB}$ is the  five dimensional  energy-momentum tensor and
$S_{AB} $ denote the energy momentum tensor confined on the brane.
We assume that the brane is located at $ y=0$ in the bulk, the
Einstein's equation on brane  is given by \cite{sasaki},
\begin{eqnarray}
\nonumber G_{AB} &=&\frac{2k_5^2}{3}[(T_{CD}-\frac{\Lambda}{k^2_5}g_{CD})~
q_A^B q_B^A+((T_{CD}-\frac{\Lambda}{k^2_5}g_{CD})n^{C}n^{D}\\
&&-\frac{1}{4}(T^{C}\;_{C}-\frac{\Lambda}{k^2_5}g_{CD}))q_{AB}]-E_{AB}+KK_{AB}-K\;^{C}_{A}K_{BC}\\
\nonumber &&-\frac{1}{2}q_{AB}(K^{2}-K^{CD}K_{CD}),
\end{eqnarray}
where $g_{AB}$ and $q_{AB}$ are related as
\begin{equation} g_{AB}=
q_{AB}+n_{A}n_{B},
\end{equation}
 and $n_A$ is a unit vector normal to the
brane.

In the present study, we assume that  bulk is empty and also the
bulk cosmological constant $\Lambda$ is considered to be absent. The
energy-momentum tensor on the brane with a zero tension  takes the
following form
\begin{equation} \label{brem}
S_{AB}=\tau_{AB}.
\end{equation}
Where $\tau_{AB}$ represents the matter fields confined on the brane
and is assumed to have a perfect fluid form. The term $E_{AB}$
appearing in the equation (3) is related to the five dimensional Weyl curvature
tensor $C_{ABCD}$ ($E_{AB} =C_{ABCD}n^Cn^D$) and $K_{AB}$ is the
extrinsic curvature associated with the brane. Assuming that bulk
space-time is $Z_2$ symmetric with respect to the brane, the Israel
Junction condition \cite{israel} implies
\begin{equation}\label{jun}
[K_{\mu\nu}]=-k^2_5(S_{\mu\nu}-\frac{1}{3}q_{\mu\nu}S).
\end{equation}
Using the condition (\ref{jun}) in (3) one obtains the effective
Einstein equation on the brane,
\begin{equation}\label{ein}
G_{\mu\nu}= k^4_5 \Pi_{\mu\nu}-E_{\mu\nu}.
\end{equation}
Where,
\begin{equation} \label{pi}
\Pi_{\mu\nu} = -\frac{1}{4}\tau_{\mu\rho}\tau^\rho_\nu+\frac{1}{12}\tau \tau_{\mu\nu}+\frac{1}{8} q_{\mu\nu} \tau_{\alpha\beta}\tau^{\alpha\beta}-\frac{1}{24}q_{\mu\nu} \tau^2.
\end{equation}
An alert reader can  immediately realise  that the  Friedmann
equation arising from equation (\ref{ein}) will be of the form $H
\propto \rho$. Thus  recovery of the standard cosmology is
impossible, which is not compatible with cosmological observations.
 This problem is due to absence of cosmological constant in bulk and tension on the brane, a point
 noted earlier by \cite{cline,kanti}. We show, in the next section, this problem can be  solved in the braneworld cosmology with a non-empty bulk.

\subsection{Effective Einstein equation with bulk scalar field }
In this section we study effective Einstein equations in presence of
the bulk scalar field and see effects in the braneworld cosmology.

We introduce  a scalar field in the bulk and cosmological constant
is still assumed to be zero, thus the corresponding five dimensional Einstein's
equation take the following form
\begin{equation}
R_{AB}-\frac{1}{2}g_{AB}R=k_5^2({T}_{AB}+S_{AB}~ \delta(y)).
\end{equation}
Where $T_{AB} $ is the energy momentum tensor of the bulk scalar
field and is given by
\begin{equation} \label{tem}
 T_{AB}= \phi,_A \phi,_B
-g_{AB}(\frac{1}{2} g^{CD}\phi,_C \phi,_D + V(\phi)).
\end{equation}
The energy momentum tensor on the brane with zero tension  is same as
(\ref{brem}) and the induced four dimensional  Einstein's equation
on brane i.e $y=\hbox {constant \,(zero)}$ hyper surface is given by equation (3).

Again assuming $Z_2$  symmetry for bulk space time \cite{israel},
using the  junction condition (\ref{jun}) and equation (\ref{tem}),
effective Einstein on the brane becomes,
\begin{equation}\label{bien}
G_{\mu\nu}=k^2_5 \hat{T}_{\mu\nu}+ K^4_5
\Pi_{\mu\nu}-E_{\mu\nu}.
\end{equation}
  Where,
\begin{equation}\label{bem}
\hat{T} _{\mu\nu}= \frac{1}{6} \left( 4 \phi,_\mu \phi,_\nu +\left(
\frac{3}{2}(\phi,_y)^2 -\frac{5}{2}q^{\alpha \beta}\phi,_\alpha
\phi,_\beta -3V(\phi)\right)q_{\mu\nu } \right),
\end{equation}
and $\Pi$ is same as  given by equation (\ref{pi}) and E$_{\mu\nu}$
is the part of the Weyl tensor.

Here also one  can see  that $ H^2 $ is not  linear in the energy density  of  brane matter.
 Since the bulk  scalar field contributes  a  term linear in its  energy density,
we can  recover standard type cosmology due to it.
\subsubsection{Friedmann equation}
The purpose of this section is  to get Friedmann like equations on
the brane. We are interested in homogeneous and isotropic geometries
on the brane, hence the metric on the brane is taken as the Friedmann-Robertson-Walker metric,
\begin{equation}\label{frwmet}
ds^{2}|_{y=0} = -dt^{2}+S^{2}(t) \delta_{ij}dx^{i}dx^{j}.
\end{equation}
The matter content on the brane is assumed to be perfect fluid form
and satisfies the usual energy-momentum conservation law. The bulk
scalar field under our consideration satisfies the following boundary condition
\begin{equation}
 \phi,y\vert_{y=0} = 0,
 \end{equation}
which means that  bulk scalar is constant with respect to $y$ on the
brane. Using equations (\ref{bien}), (\ref{bem}) and
(\ref{frwmet}), the Friedmann equation on the brane is obtained  as,
\begin{equation} \label{flat}
\left(\frac{\dot{S}}{S}\right)^2 + \frac{k}{S^2}  = \frac{k^2_5}{3}
\rho_B + \frac{k^4_5}{36} \rho^2_b -\frac{E_{00}}{3}.
\end{equation}
 Where  $\rho_b$ is the brane energy density and $\rho_B$ is the bulk energy density and can be obtained from (\ref{bem})  as
\begin{equation} \label{energy}
\rho_B = \frac{1}{2}\left(\frac{1}{2}\dot{\phi}^2 + V(\phi)\right).
\end{equation}
 Next, consider a model in which the brane field  and dark
radiation terms ( $ E_{00}$ ) on the right side of  Friedmann's equation are
negligible, then  Eqn \ref{flat} becomes
\begin{equation}\label{frw}
H^2 + \frac{k}{S^2} = \frac{k^2_5}{3} \rho_B,
\end{equation}
 which implies that dynamics on the brane are governed by the bulk scalar field.

Next, we are interested in  the  inflationary phase  driven by  the
bulk scalar field rather than due to inflaton on the brane. Consider
a universe where the bulk scalar field dominates, the equation
governing the scalar field is the
 Klein-Gordon equation. Using the full five dimensional metric (\ref{met})
 with a Friedmann- Robertson-Walker metric for brane part,  the Klein-Gordon equation is obtained as,
 \begin{equation}
 \ddot{\phi} + 3\left(\frac{
\dot{S}}{S}\right ) \dot{\phi}
 -\partial^2_y \phi+  V^{\prime}= 0.
 \end{equation}
 In view of the slow-roll approximation the Klein-Gordon
equation  on the brane becomes,
 \begin{equation}\label{em}
 3 H \dot{\phi}+ V^{\prime} =0.
 \end{equation}
 The condition for acceleration  can be found from (\ref{frw}),  by assuming that the  bulk scalar field follow the usual
 energy  momentum conservation law for the bulk matter  on the brane i.e, $\dot{\rho_B}= -3H(P_B+ \rho_B) $, as:
\begin{equation}
\frac{\ddot{S}}{S} =  -\frac{k^2_5}{6} \left( \rho_B + 3  P_B
\right)> 0
\end{equation}
from this one obtains,
 \begin{equation}\label{accl}
  P_B <  -\frac{\rho_{B}}{3}
\end{equation}
which matches the acceleration conditions of the standard cosmology.

Our next goal is to obtain the acceleration condition  for a universe  with scalar field in the bulk.
For this, the energy density of bulk scalar field can be considered as the form given
by equation (\ref{energy}) and  pressure is obtained from
(\ref{bem}) as $P_B = \frac{5}{2}\dot{\phi}^2 - 3 V $. Thus, in the
scalar field dominated universe the acceleration condition can be
derived using equation (\ref{accl}) as
\begin{equation}\label{slowrol}
\dot{\phi}^2 < \frac{2}{3}V.
\end{equation}
Which  is similar to the standard acceleration condition except for a factor
difference. This  indicate that the bulk scalar field need more potential energy than the standard inflaton field for inflation to occur.
\section{Bulk scalar field in DGP braneworld }
In  the  DGP model our universe a is a three brane embedded in five dimensional  bulk
and there is an induced four dimensional Ricci scalar on the brane, due to
radiative correction to the graviton propagator on the brane. In
this model there is a length scale below which the  potential has
usual Newtonian form and above which the gravity becomes five
dimensional. The cross over scale between the four dimensional and
five dimensional gravity is
 \begin{equation}
 r_c=\frac{k_5^2}{2\mu^2}.
 \end{equation}
Where $\mu^2= 8 \pi G_{4}$.
 We start with a generalised DGP model in which both bulk cosmological
 constant $\Lambda$ and brane tension $\sigma$  are
non-zero. In our model we consider a non-empty bulk, with a bulk
scalar field living in it.  The bulk energy momentum tensor is given
by equation (\ref{tem}) and the four dimensional energy-momentum tensor given by
\begin{equation}
S_{AB}=\tau_{AB}-\sigma q_{AB} -\mu^{-2} G_{AB}.
\end{equation}
Where last term in the above equation represents the induced term.
Again assuming bulk is $Z_2$ symmetric, the Einstein equation on the brane is obtained  as:
\begin{eqnarray}\label{gendgp}
 \left( 1+\frac{\sigma k^2_5}{6 \mu^2}\right )
G_{\mu\nu}&=&-\left(\frac{k^2_5 \Lambda}{2}+\frac{k^4_5\sigma^2}{12}\right
)q_{\mu\nu}+\mu^2 \tilde{T}_{\mu\nu}  \nonumber  \\
&& +\frac{\sigma k^4_5}{6}\tau_{\mu\nu} +\frac{k^4_5}{\mu^4} F_{\mu\nu}
+ k^4_5 \Pi_{\mu\nu} + \frac{ k^4_5}{\mu^2} L_{\mu\nu}-E_{\mu\nu},
\end{eqnarray}
where $\tilde{T}_{\mu\nu}$,
 $\Pi_{\mu\nu} , F_{\mu\nu} \hbox {\, and } \, L_{\mu\nu}$
are respectively given by,
\begin{eqnarray}\label{rbem}
\tilde{T} _{\mu\nu}&=&  \frac{r_c}{3} \left[  \left( 4 \phi,_\mu
\phi,_\nu +\left( \frac{3}{2}(\phi,_\chi)^2 -\frac{5}{2}q^{\alpha
\beta}\phi,_\alpha \phi,_\beta -3V(\phi)\right)q_{\mu\nu }
\right)\right],
\\
\Pi_{\mu\nu}& =& -\frac{1}{4}\tau_{\mu\rho}\tau^\rho_\nu+\frac{1}{12}\tau \tau_{\mu\nu}+\frac{1}{8} q_{\mu\nu} \tau_{\alpha\beta}\tau^{\alpha\beta}-\frac{1}{24}q_{\mu\nu} \tau^2,
\\
F_{\mu\nu} &=& -\frac{1}{4}G_{\mu\rho}G^\rho_\nu+\frac{1}{12} G
G_{\mu\nu}+\frac{1}{8} q_{\mu\nu}
G_{\alpha\beta}G^{\alpha\beta}-\frac{1}{24}q_{\mu\nu} G^2,
\\
L_{\mu\nu} &=& \frac{1}{4}(G_{\mu\rho}\tau^\rho_\nu + \tau_{\mu\rho} G^\rho_\nu ) - \frac{1}{12} (\tau G_{\mu\nu} + G \tau_{\mu\nu} )-\frac{1}{4} q_{\mu\nu} (G_{\alpha\beta}\tau^{\alpha\beta}-\frac{1}{3} G \tau).
\end{eqnarray}
As mentioned earlier $E_{\mu\nu}$ is the projection of the Weyl
tensor on the brane. The  Einstein equation in DGP cosmology is
differrent from equations (\ref{ein}, \ref{bien}) as it contains
extra term, which is quadratic in $G_{\mu\nu}$ due to inclusion of
$G_{\mu\nu}$ in the four dimensional energy momentum tensor.

 At this stage we switch to the original DGP model where cosmological constant
and brane tension are zero then  equation (\ref{gendgp}) becomes,
\begin{equation}\label{dgpe}
G_{\mu\nu}=\mu^2 \tilde{T}_{\mu\nu}+ \frac{k^4_5}{\mu^4} F_{\mu\nu} +
k^4_5  \Pi_{\mu\nu} + \frac{ k^4_5}{\mu^2} L_{\mu\nu}-E_{\mu\nu}.
\end{equation}
\subsection{The  Friedmann's equation in DGP cosmology}
In this section we  discuss the  Friedmann type equations, for this the
universe is considered homogeneous and isotropic on the brane. The
metric on the brane is  the  Friedmann-Robertson-Walker metric
(\ref{frwmet}), taking the 00 component of  the  equation
(\ref{dgpe}) becomes
\begin{equation}
 G_{00} = \mu^2\rho_B + \frac{k^4_5}{12}
\rho^2_b +\frac{k^4_5}{12 \mu^4} G_{00}^2 - \frac{k^4_5}{6\mu^2}
\rho_b G_{00}-\frac{\cal{E} }{S^4}\nonumber
\end{equation}
 which can be rewritten as
 \begin{equation} \label{dgp} \left(H^2+\frac{k}{S^2}\right)= \epsilon \frac{2
\mu^2}{k^2_5}\sqrt{\left(H^2+\frac{k}{S^2}\right)-\frac{\mu^2}{3}\rho_B-\frac{\cal{E}}{S^4}} +
\frac{\mu^2}{3} \rho_b.
\end{equation}
 This  is the general Friedmann equation in DGP cosmology and it matches with the result obtained in \cite{def}. The
 $\epsilon=\pm 1 $ corresponds to two possible embedding  of the brane in
the bulk. Consider a case in which $\mu$ go to inifty i.e induced
curvature term is not there  ( and  $\cal{E}$  is negligible  ) then
equation (\ref{dgp}) becomes
\begin{equation}
 H^2+\frac{k}{S^2}= \frac{\mu^2}{3}
\rho_B + \frac{k^4_5}{36} \rho^2_b,
\end{equation}
 which is the Friedmann equation  obtained  in (\ref{flat}). Hence when the curvature term is not included in $S_{AB}$ (or in  action)
 our result reduces to the flat case and is same as in \cite{binetury,sasaki}.

\subsection{Standard cosmology with a bulk scalar field}
 The  standard cosmology is recovered from  DGP braneworld scenario, with a brane matter by applying suitable
 conditions \cite{def}. However we are interested in seeing the dynamics of brane due to a bulk scalar
 field and to obtain standard type cosmology with bulk field.
We  assume  that brane has no matter and also  $\cal{E}$ is zero,
then equation (\ref{dgp}) becomes,
\begin{equation} \label{ndgp}
\left(H^2+\frac{k}{S^2}\right)\frac{k^2_5}{2 \mu^2} -\epsilon
\sqrt{\left(H^2+\frac{k}{S^2}\right)-\frac{\mu^2}{3}\rho_B}=0.
 \end{equation}
 We  get standard type cosmology from the above equation by imposing the
 condition that first term in (\ref{ndgp}) should be negligible
i.e,
\begin{equation}
\left(H^2+\frac{k}{S^2}\right) {\left( \frac{k^2_5}{2 \mu^2} \right )}^2  \ll 1 \;
\end{equation}
thus,
\begin{equation}\label{stddgp}
 H^2+\frac{k}{S^2}= \frac{\mu^2}{3} \rho_B.
\end{equation}
Note that the  condition (35) for recovery of standard
cosmology is different  from one obtained in \cite{def} with  matter
field on the brane.
The bulk energy density  is given by,
\begin{eqnarray} \rho_B&=&r_c(\frac{1}{2}\dot{\phi}^2 + V(\phi)) \nonumber \\
& =&\frac{1}{2}\dot{\Phi}^2 + \tilde{V}(\Phi).
 \end{eqnarray}
Where we introduced  the effective rescaled field $\Phi$ on the
brane, \begin{equation}
\Phi=\sqrt{r_c}\phi
\end{equation}
 and the potential is rescaled as,
\begin{equation}
 \tilde{V}(\Phi)= r_c V(\frac{\Phi}{\sqrt{r_c}}).
\end{equation}
 The
interesting feature of the  effective field  $\Phi$ is that, the rescaling
depends on the cross over scale. Our study of the rescaling the scalar field and hence the potential is  without taking a specific form. However one can consider different forms of the potential and check that the resclaing holds. The rescaled bulk scalar field on the brane act as an effective field and thus the dynamics of the brane can be goverened  by it. For a universe dominated by a bulk scalar field, the condition for acceleration in terms of the  rescaled field, on the brane is
obtained as,
\begin{equation}
\dot{\Phi}^2 < \frac{2}{3}\tilde{V}.
\end{equation}
As mentioned earlier the only modification coming is a factor
compared to the standard inflation.

In view of the rescaling one can immediately infer that,
following the relation (38), the cross overscale  can be  written in the present context as
\begin{equation}
  r_c = \frac{\Phi^2}{\phi^2}.
\end{equation}
\subsection{Late time cosmology and solutions to Friedmann equation}
The most important aspect of DGP model is the self accelerating
solutions in the late universe. Let us  see, the presence of bulk
scalar field  can give rise self acceleration in DGP model. For
this, consider the two branches of solution of the effective Friedmann equation (\ref{dgp}) which
depend upon the value of $\epsilon$ and
 in absence of $\rho_b $ it can be rewritten  as,
\begin{eqnarray} \label{self}
H^2+\frac{k}{S^2}= \frac{1}{2r_c^2}\left[ 1 + \epsilon
\sqrt{1-\frac{4 \mu^2}{3}\rho_B r_c^2  }\right].
\end{eqnarray}
As mentioned earlier, the bulk scalar field on the brane satisfies the usual  energy momentum consevation law and hence it follow that $ \rho_{B} \propto S^{-3}$. Thus
equation (\ref{self}) can be expanded under the condition that  $\mu^2 \rho_B \ll1/r^2_c$.
At zeroth order for $k=0$ case, the two branches  are
\\
$\bullet$  $\epsilon=-1$
\begin{equation}
H^2= 0
\end{equation}
correspond to a  Friedmann universe  with  $ S \sim \hbox{constant} $, which means that it is asymptotically static.
\\
~~~~~~$\bullet$  $\epsilon=1$ \\
\begin{equation}
 H^2= \frac{1}{r_c^2}
 \end{equation}
 which is the self-accelerated solution in agreement with
\cite{def}, and   $ S \sim \hbox{exp}(t/r_c)$, which shows the
self-accelerating phase of the universe. Therefore, we can consider a model of the universe that filled with  scalar field in bulk   and can lead to a self accelerating phase of the universe as in the case of the cosmolgical models with inflaton or matter field in the brane. Thus the late time behaviour of the universe does not alter even if one start a cosmological model with  scalar field living in the bulk without  matter content in the brane.
 Hene the effective rescaled bulk scalar field  may be an alernative to the  inflaton on the brane.
\section{Conclusions}
We examined  the effects of bulk scalar field in  the braneworld cosmological
scenario. Our basic set up is  a three dimensional brane embedded in a five dimensional space-time with an infinite size extra-dimension.
 We  derived the effective Einstein equations on the brane  for different cases.

It is noted that
the effective Einstein equation,
 in the  case of zero tension brane and cosmological constant, leads to
Friedmann equation with  $H \propto \rho$ rather than usual situation where $H \propto
\sqrt{\rho}$. Therefore the aformetioned case
cannot lead  to the standard type of  cosmology, which confronts with cosmological observations so far.
In order to overcome this unlikely situation, one has to
introduce  a bulk cosmological constant and brane tension in a braneworld cosmological model under the consideration.
As an alternative, we show that the  introduction of
 scalar field in the bulk can evade this problem. We have shown that  the  derived
Einstein's equation  can recover  the standard Friedmann equation with a bulk scalar field in absence of the  brane tension and cosmological constant. Using this model  inflation  driven by the  bulk
scalar field  on the brane is also examined and
the  derived conditions for acceleration are  matching with the standard case except for a factor difference. This is due the modification coming from the extra-dimensional effect. It further indicates that the required potential energy due to the effective bulk scalar field is much higher than that of the inflaton on the brane for inflation to occur.

The Einstein equation in DGP cosmology are also derived, then we
recovered the  standard type of cosmology in terms of the  bulk scalar field. The
condition for recovery of standard cosmology is different from the
one obtained with a brane field. And it shows that condition is
reversed in comparision with \cite{def}, interestingly  still H goes as $r_c^{-1}$.
We introduced a rescaled bulk scalar field on the brane in the DGP model and the rescaling
depends on scale factor $r_c$. Though we have introduced the rescaling of the potential in a general manner,  the validity  of the rescaling procedure can be examined by taking  specific forms of the potential.
 The DGP model with the rescaled bulk scalar field on the brane can also provide a self  accelerating  solution.
Thus our results has concordance with Deffayet's result \cite{def} in the late-time behaviour of the  universe. Therefore the rescaled bulk scalar field  evaluated on the brane  can
indeed replace the brane field and can govern the  dynamics on the brane.

The  rescaling procedure of the field helps us to express the crossover scale  as the ratio  of the effective scalar field to the  bulk  field. Although we have considered brane as empty in this work, it would be interesting to  retain the  brane field  and consider a possible bulk-brane fields interactions. One could then examine whether this system  address the issue of alternative to dark energy and thus test the viability of this model with respect to observations.

\end{document}